\documentclass[aip,jcp,preprint,noshowkeys,superscriptaddress]{revtex4-2}

\usepackage{graphicx,bm,xcolor,microtype,multirow,amscd,amsmath,amssymb,amsfonts,physics,longtable,wrapfig,txfonts,soul}
\usepackage[normalem]{ulem}

\usepackage[utf8]{inputenc}
\usepackage[T1]{fontenc}
\usepackage{txfonts}

\usepackage[
	colorlinks=true,
    citecolor=blue,
    breaklinks=true
	]{hyperref}
\usepackage{}
\DeclareMathOperator*{\argmin}{argmin} 

\usepackage{scalerel}

\newcommand{\rep}[1]{\textcolor{black}{#1}}
\newcommand{\strikeout}[1]{}

\makeatletter
\def\@email#1#2{%
 \endgroup
 \patchcmd{\titleblock@produce}
  {\frontmatter@RRAPformat}
  {\frontmatter@RRAPformat{\produce@RRAP{*#1\href{mailto:#2}{#2}}}\frontmatter@RRAPformat}
  {}{}
}%
\makeatother
\DeclareUnicodeCharacter{2212}{-}

\begin{document}	


\title{ 
 Static versus dynamically polarizable environments within the many-body $\bf{GW}$ formalism  \rep{ \strikeout{Assessing the antiadiabatic limit}}   }

\author{David Amblard}
\affiliation{Univ. Grenoble Alpes, CNRS, Inst NEEL, F-38042 Grenoble, France}

\author{Xavier Blase}
\affiliation{Univ. Grenoble Alpes, CNRS, Inst NEEL, F-38042 Grenoble, France}
\email{xavier.blase@neel.cnrs.fr}

\author{Ivan Duchemin}
\affiliation{Univ. Grenoble Alpes, CEA, IRIG-MEM-L\_Sim, 38054 Grenoble, France}
\email{ivan.duchemin@cea.fr}

\date{\today}

\begin{abstract}
Continuum or discrete polarizable models for the study of optoelectronic processes in embedded subsystems  rely mostly on the restriction of the surrounding electronic dielectric response  to its low frequency limit. Such a description hinges on the assumption that the electrons in the surrounding medium react instantaneously to any excitation in the central subsystem, \strikeout{which is thus treated in the antiadiabatic limit}
\rep{treating thus the environment in the adiabatic limit}. Exploiting a recently developed embedded $GW$ formalism, with an environment described at the fully \textit{ab initio} level, we assess the merits of the \rep{\strikeout{anti}} adiabatic limit with respect to an environment where the full dynamics of the dielectric response is considered. Further, we show how to properly take the static limit of the environment susceptibility, introducing the so-called Coulomb-hole and screened-exchange contributions to the reaction field. As a first application, we consider  a C$_{60}$ molecule at the surface of a C$_{60}$   crystal, namely a case where the dynamics of the embedded and embedding subsystems  are similar. The common \rep{\strikeout{anti}}adiabatic assumption, when properly treated, generates errors below  $10\%$ on the polarization energy associated with frontier energy levels and associated energy gaps.
Finally, we consider a water molecule inside a metallic nanotube, \rep{\strikeout{the worst situation for the antiadiabatic limit}} \rep{the worst case for the environment adiabatic limit. The error on the gap polarization energy remains below 10$\%$, even though the error on the frontier orbitals polarization energies can reach a few tenths of an electronvolt.  } 
\end{abstract}


\maketitle

\section{Introduction}
\label{sec:intro}

The description of the electronic  properties of  a quantum subsystem  embedded in a large and complex electrostatic and dielectric environment stands as a severe challenge to quantum mechanical modeling. Such an environment can be very large, disordered,  dynamical in the case of molecular solvation, so that brute force quantum mechanical treatments may prove out-of-reach. Still, in many applications pertaining to organic optoelectronics, wet chemistry, biology, etc., the surrounding medium cannot be ignored, driving significant shifts of the electronic energy levels and optical excitations. Such a situation spawned hybrid strategies, merging the quantum mechanical treatment of the ``{a}ctive'' subsystem, with a simplified description of the environment as an effective continuum or atomistic medium. 
\cite{WARSHEL1976,Miertus_1981}


Concerning the quantum mechanical framework used to describe the embedded subsystem, a specific formulation of Green's function many-body perturbation theory, the $GW$  formalism, \cite{Hed65} has gained much attention in recent years in the quantum chemistry community, offering a favorable compromise between accuracy and computational efficiency. The $GW$  approach  tackles the calculation of electronic energy levels, as properly defined by a photo-emission experiment. Numerous benchmarks for gas phase molecular systems, with in most recent studies comparisons to higher-level coupled-cluster reference calculations, are now available. \cite{Rostgaard2010,Blase2011,Marom2012,Bruneval2013,vanSetten2015,Krause2015,Knight2016,Caruso2016,Rangel2016,Kaplan2016,Maggio2017,Govoni2018,Jin2019,Weiwei2019,Forster2021,Jiachen2022}

A central feature of such a formalism  is that the    correlation potential, labeled a self-energy, relies on the knowledge of the dynamically screened Coulomb potential $W({\bf r},{\bf r}'; \omega)$. As such, environmental dielectric screening effects can be rather straightforwardly combined with the $GW$ framework. Further, the dynamical nature of the self-energy allows exploring the effect of dynamical screening, namely the fact that the environment dielectric function $\epsilon({\bf r},{\bf r}'; \omega)$ or electronic susceptibility $\chi({\bf r},{\bf r}'; \omega)$ are frequency-dependent. 

Together with the integration of the $GW$ formalism with polarizable continuous models (PCM), \cite{Duchemin2016,Duchemin2018,Clary2023} 
the merging of the $GW$ formalism with semi-empirical discrete polarizable approaches for the environment, extending the general family of hybrid quantum/classical (QM/MMpol) techniques for complex polarizable environments,  was explored for the study of embedded molecular systems. \cite{Baumeier2014,Varsano2014,Duchemin2016,Li2016,Li2018,Wehner2018,Tirimbo2020} 
In such approaches, the dielectric response from the environment relies mainly on effective atomic polarizabilities \cite{Thole1981} taken in their low frequency limit
$\alpha_{\mathrm{at}}( \omega \rightarrow 0)$ (for a recent overview, see e.g. Refs.~\citenum{D_Avino_2016,mennucci-MMpol-2020}). Responding to an electronic excitation in the central embedded subsystem, atoms in the environment develop induced dipoles that  generate a reaction field stabilizing  the  electronic excitation $\varepsilon_n$. The associated energy level shifts $P_n = \varepsilon_n^\text{\,embedded} - \varepsilon_n^\text{\,gas}$, from the isolated (gas) to the embedded phase, are labeled polarization energies and can be as large as several electronvolts. 

Taking the environment electronic susceptibility $\chi(\omega)$ in its low frequency limit ($\omega \rightarrow 0$) hinges on the assumption that the electrons in the environment respond instantaneously to an electronic excitation in the central subsystem. This is an adiabatic approximation for the environment \rep{\strikeout{, or equivalently, an antiadiabatic approximation for the embedded subsystem}}. Such an approximation is expected to be   valid   if the  gap of the environment is much larger than that of the central subsystem. In other situations, where such a decoupling is not verified, in particular when the environment is characterized by a gap smaller than that of the embedded subsystem, such approximations needs to be validated. A critical assessment was recently proposed on the basis of model  solutes and solvents. \cite{Painelli2020} Also, a pioneering dynamical implementation of the PCM combined with the $GW$ formalism was proposed, but without  comparisons to results obtained in the static PCM limit.  \cite{Clary2023}   Similar questions have been raised in the different context of Dynamical Mean Field Theory. The onsite Coulomb integrals $U$ between strongly-correlated electrons are screened by the weakly-correlated electronic degrees of freedom, with a choice to make between a static $U(\omega \to 0)$ and a dynamical $U(\omega)$ that is much more difficult to handle. \cite{Aryasetiawan2004,Casula2012}

Recently, a fully \textit{ab initio} QM/QM' $GW$ formalism was developed, allowing to consider very large atomistic environments with a dielectric response described at the random phase approximation (RPA) level. \cite{Amblard2023}  Initially treated in the static limit, the \textit{ab initio} treatment of the  RPA dielectric response allows  exploring straightforwardly the impact of switching the full dynamics of the environment susceptibility. This is what we target in the present study. We explore some common situations where \rep{\strikeout{the antiadiabatic approximation for the embedded subsystem}} \rep{the adiabatic approximation for the environment}  is expected to fail. We consider in particular the case of a fullerene at the surface of a fullerene crystal, a situation where there is no energy decoupling between the embedded and embedding subsystems electronic response. We show however that  when properly implemented, the adiabatic approximation for the surrounding polarizable medium does not induce errors larger than 10$\%$ for the polarization energies. Further, we explore the case of a water molecule inside a metallic nanotube, \rep{\strikeout{where the antiadiabatic limit should be applied to the environment (the nanotube) and not to the central water molecule}} \rep{where the adiabatic approximation should be used for the large gap water molecule, not for the environment (the metallic nanotube)}.  \rep{ The error associated with the energy gap remains below 10$\%$, but with errors as large as a few tenths of an eV for individual frontier orbitals. } We compare the accuracy of various implementations of the adiabatic approximation for the environment within the $GW$ framework. Our results suggest to favor a strategy where the central subsystem is treated at the fully dynamical $GW$ level, while the environment static reaction field is treated within the analog of the static Coulomb-hole plus screened-exchange approximation.   

\section{Theory }

In the following, we briefly describe the salient features of the $GW$ formalism and its merging with a polarizable  environment, including the specific aspects associated with the definition and construction of the reaction field. Extended details about the $GW$ formalism can be found in the literature, from seminal articles \cite{Hed65,Str80,Hyb86,God88,Lin88,Farid88} to recent reviews or books. \cite{Ary98,Farid99,Oni02,Pin13,Leng2016,ReiningBook,Gol19rev}

\subsection{The $GW$ self-energy }

The $GW$ formalism is a specific Green's function many-body perturbation theory taking the one-body time-ordered Green's function $G({\bf r},{\bf r}'; \omega)$ as a central variable, in place e.g. of the charge density in Density Functional Theory (DFT). Rather than performing a perturbation theory in terms of the bare Coulomb potential $v$, the $GW$ exchange-correlation self-energy can be understood as a first-order expansion in the screened Coulomb potential $W$:
\begin{align}
 \Sigma({\bf r},{\bf r}'; E) = \frac{i}{2\pi} \int \dd\omega \; \mathrm{e}^{i \eta \omega} G({\bf r},{\bf r}'; E+\omega)  W({\bf r},{\bf r}'; \omega)
 \label{eqn:sigma}
\end{align}
with $\eta$ a positive infinitesimal. Contrary to standard DFT exchange-correlation functionals, the self-energy is dynamical. Performing $GW$ calculations in a polarizable or dielectric environment relies straightforwardly on the relation between $W({\bf r},{\bf r}'; \omega)$ and the 
electronic susceptibility ${\chi}({\bf r},{\bf r}'; \omega)$ :
\begin{equation}
\begin{split}
    W({\bf r},{\bf r}'; \omega)   &= v({\bf r},{\bf r}') +  \\
       &\int \dd{\bf r}_1 \dd{\bf r}_2 v({\bf r},{\bf r}_1) {\chi}({\bf r}_1,{\bf r}_2; \omega) v({\bf r}_2,{\bf r}' ),  
     \label{eqn:W_chi}
\end{split}
\end{equation}
with $\chi$ related to the non-interacting susceptibility $\chi_0$:
\begin{equation}
\begin{split}
    \chi({\bf r},{\bf r}'; \omega) & = \chi_0({\bf r},{\bf r}';\omega) +\\  &\int \dd{\bf r}_1 \dd{\bf r}_2
      \chi_0({\bf r},{\bf r}_1;\omega) v({\bf r}_1,{\bf r}_2) {\chi}({\bf r}_2,{\bf r}'; \omega), \label{eqn:dysonchi}
\end{split}
\end{equation}
where the latter equation holds within the random phase approximation (RPA). 
Starting traditionally from a  Kohn-Sham or Hartree-Fock calculation, the input Green's function and susceptibility are built from the resulting mean-field one-body eigenstates:
\begin{align}
    G({\bf r},{\bf r}'; \omega) &= \sum_n \frac{ \phi_n({\bf r}) \phi_n^{*}({\bf r}') }{\omega - \varepsilon_n +   i \eta \times \text{sgn}(\varepsilon_n - \mu) } \\
    \chi_{0} ({\bf r},{\bf r}'; \omega) &=   \sum_{n,m}  \frac{ (f_m -f_n) \; \phi_n({\bf r}) \phi_m^{*}({\bf r})  \phi_n^{*}({\bf r}') \phi_m({\bf r}') }
    { \omega - ( \varepsilon_n - \varepsilon_m) + i \eta \times \text{sgn}( \varepsilon_n - \varepsilon_m ) }, \label{eqn:xi0}
\end{align}
where $\{f_{n/m}\}$ are occupation numbers and $\mu$  the zero-temperature chemical potential. The analysis of the independent-electron susceptibility, the main computational bottleneck, shows that it can be built with an $\mathcal{O}(N^4)$ computational effort with respect to the number of electrons $N$, but  various space-time,  interpolative separable density fitting, stochastic or moment-conserving reformulations can dramatically reduce this computational complexity.  \cite{Rojas1995,Foerster2011,Neuhauser2014,Liu2016,Vlcek2017,Vlcek2018,Wilhelm2018,Wilhelm2018,Forster2020,Kim2020,Kupetov2020,Gao2020,Wilhelm2021,Duchemin2021,Forster2023,Scott2023}
The construction of the Green's function and screened-Coulomb potential with input Hartree-Fock, Kohn-Sham or hybrid exchange-correlation functionals (XCF) is labelled the single-shot $G_0W_0$@XCF scheme and the impact of the choice of the XCF has been extensively studied. 
\cite{Blase2011,Bruneval2013,Rangel2016,Gui2018,Kshirsagar2021,Kshirsagar2023}
Various self-consistent schemes, reinjecting only the corrected energy levels, \cite{Shishkin2007,Blase2011,Kaplan2016} 
and further the updated eigenfunctions as well in a fully self-consistent scheme, \cite{Schilfgaarde2006,Stan2006,Rostgaard2010,Marom2012,Koval2014,Kaplan2016,Knight2016,Marie2023} allow reducing the impact of the choice of the initial XCF, curing further problems associated with multiple quasiparticle solutions. \cite{Veril2018,Monino2022} Core level spectroscopies \cite{Golze2018,Kehry2020,Keller2020,Mejia2022,Li2022,Yao2022,Kahk2023,Mukatayev2023} and inclusion of relativistic effects \cite{Umari2014,Scherpelz2016,Holzer2019,Kehry2020,Keller2020,Yeh2022} are currently the focus of a growing number of studies. 
Comparisons with coupled-cluster techniques have been pioneered \cite{Lange2018,Quintero2022,Tolle2023} and the robustness of the $GW$ approximation applied to strongly correlated  or multi-reference systems, including Hubbard models, is a current subject of investigation.  \cite{Verdozzi1995,Sabatino2021,Honet2022,Ammar2024}

In the present study, concerned with comparing static versus dynamical $GW$ calculations, we will be considering  further  the so-called  Coulomb-hole (COH) plus screened-exchange (SEX) static limit\cite{Hed65,Hyb86,Bruneval2005,Bruneval2006,Berger2021} to the $GW$ self-energy, with:
\begin{subequations}
\begin{align}
\Sigma^{\text{SEX}}({\bf r},{\bf r}') &= - \sum_i^{\text{occp}}  \phi_i({\bf r}) \,\phi^{*}_i({\bf r}') \, W({\bf r},{\bf r}';\omega=0) \label{SEX} \\  
\Sigma^{\text{COH}}({\bf r},{\bf r}') &= \frac{1}{2} \,\sum_n   \phi_n({\bf r}) \,\phi^{*}_n({\bf r}') \, \left[ W({\bf r},{\bf r}';\omega=0)-v({\bf r},{\bf r}') \right], \label{COH}
\end{align}
\end{subequations}
where the index $(i)$ runs over the occupied states. In the static COHSEX approach, only the low-frequency $W(\omega \rightarrow 0)$ screened Coulomb potential is required.  An important observation is that taking abruptly $\omega$ to zero for the screened Coulomb potential $W(\omega)$ in Eq.~\eqref{eqn:sigma} just yields the SEX term, neglecting the COH term that stems from the poles of the dynamically screened Coulomb potential. As analyzed first in Ref.~\citenum{Bruneval2006}, the use of the full static COHSEX Hamiltonian provides a much better approximation to the $GW$ operator as compared to the static screened-exchange-only (SEX) term. 

An elegant and simple way to recover the full COHSEX approximation, while having as an information only the low-frequency limit of the screened Coulomb potential $W(\omega \rightarrow 0)$, or equivalently of the susceptibility $\chi(\omega \rightarrow 0)$, is to assume a  simple pole model, namely: 
\begin{align}\label{eqn:ximodel}
\begin{split}
    \chi_\lambda({\bf r},{\bf r}'; \omega) &= \chi({\bf r},{\bf r}'; \omega=0)  \\   
          &\times \frac{\lambda}{2}\left[  \frac{1}{\omega + \lambda-i\eta}- \frac{1}{\omega - \lambda+i\eta}  \right],
\end{split}
\end{align}
with ($\lambda$) a unique pole energy for simplicity. Such an expression has the correct low and high frequency limits and time-ordering structure in the energy plane. As shown in the Appendix \ref{part:COHSEX_pole_model}, the full static COHSEX expression can be simply recovered by taking the pole energy ($\lambda$) to infinity \textit{after} performing the energy integration.  The same strategy will be used here below in order to merge properly a polarizable environment, restricted to its low-frequency  susceptibility limit, with a fully dynamical $GW$ formalism for the central subsystem.

\subsection{Embedding}

Our embedding strategy relies on the fragment, or subsystem, approximation, assuming a weak overlap between the wavefunctions of the various fragments. 
Such an approximation was already  implemented at the many-body $GW$ and Bethe-Salpeter equation (BSE) levels,  not only in the ideal case of weakly interacting molecular systems, \cite{Fujita2018,Fujita2019,Fujita2021,Tolle2021,Amblard2023,Weng2023} 
but also for physisorbed molecules on surfaces, \cite{Neaton2006,Liu2019,Liu2020,Xuan2019,Cheng2021}  or weakly bonded layered 2D materials. \cite{Andersen2015,Winther2017,Amblard2022,Despoja2024} It was further recently shown that the fragmentation in subdomains of an  hexagonal boron-nitride sheet could be achieved without altering its short to long-range dielectric properties, with application to the study of defects in the true dilute limit. \cite{Amblard2022}   In the present case of molecular systems, the application of the fragment approximation in the  construction of the environment susceptibility was found to affect the polarization energy by an error no larger than a percent (see Supplementary Material Ref.~\citenum{Amblard2023}). Strategies to go beyond the fragment approximation have been developed at the ground-state DFT \cite{Jacob2014} and many-body \cite{Fujita2021,Cheng2021} levels, attempting in particular to correct the non-interacting susceptibility by the   effect  of the residual interaction between subsystems.   Such developments stand aside of the present work devoted to the validity of the adiabatic approximation for the environment electronic response. 

\subsubsection{ Generalities }

Partitioning the system into the central subsystem of interest (labeled by 1) and the environment (labeled by 2), and assuming non-overlapping wavefunctions between subsystems (1) and (2), results in a block diagonal $\chi_0$ operator and leads to rewriting the screened Coulomb potential as follows:
\begin{align}
    W^{-1}(\omega) &= v^{-1} - \chi_0^{(1)}(\omega) - \chi_0^{(2)}(\omega) \\
           &= {\tilde v}^{-1} - \chi_0^{(1)}(\omega) \\
    {\tilde v}(\omega) &= v + v \; \chi_0^{(2)}(\omega) \; {\tilde v}(\omega),
\end{align}
where we dropped the position variables. The restriction $W_{11}$ of the screened Coulomb potential to the solute then reads, using block notations:
\begin{align}
  W_{11}(\omega) &= {\tilde v}_{11}(\omega) + {\tilde v}_{11} \chi_0^{(1)}
   W_{11}(\omega) \label{W_11}\\
   {\tilde v}_{11}(\omega)
   &= v_{11} + v_{12} \; \chi_0^{(2)}(\omega) \; {\tilde v}_{21}(\omega),
\end{align}
where e.g. $v_{12}$ is the bare Coulomb potential connecting points in  the central subsystem (1) to points in the environment (2). The latter equation can be rewritten:
\begin{align}
    {\tilde v}_{11}(\omega)
   &= v_{11} + v_{\text{reac}}(\omega) \\
   v_{\text{reac}}(\omega) &= v_{12} \; \chi^{(2)}(\omega) \;  v_{21}
   \label{eqn:vreac}
\end{align}
where $\chi^{(2)}(\omega)$ is the interacting susceptibility of the environment \textit{in the absence of the solute.} The potential $v_{\text{reac}}$ is the so-called reaction field: qualitatively, a charge density change in the subsystem (1) upon excitation generates through the $v_{21}$ Coulomb interaction a charge density change in the environment (2) proportional to $\chi^{(2)}v_{21}$ that in return  exerts a reaction field on (1) via $v_{12}$.  Finally, the Eq.~\eqref{W_11} can be rewritten
\begin{align}
    W_{11}(\omega)&={\tilde v }_{11}(\omega)+{\tilde v }_{11}(\omega)\cdot{\tilde \chi^{(1)}}(\omega)\cdot{\tilde v }_{11}(\omega) \\
    {\tilde \chi^{(1)}}(\omega)&= \chi_0^{(1)}(\omega)+ \chi_0^{(1)}(\omega)\cdot{\tilde v }_{11}(\omega)\cdot{\tilde \chi^{(1)}}(\omega). \label{eqn:tildexi}
\end{align}
${\tilde \chi^{(1)}}(\omega)$ appears as an interacting susceptibility of the central subsystem of interest (1) when its Coulomb interactions are renormalized by the reaction field created by the environment (2).  

\subsubsection{ The ${\Delta}\mathrm{COHSEX}$ approach }

In  several studies merging the $GW$ formalism  with semi-empirical or fully \textit{ab initio} polarizable environments, the polarization energy was efficiently calculated as an energy difference between two static COHSEX calculations performed with and without the environment, \cite{Neaton2006,Duchemin2016,Li2016,Li2018,Liu2019,Amblard2022,Amblard2023} namely:
\begin{align}
    P_n^{\Delta \mathrm{COHSEX}} = \varepsilon_n^{\text{COHSEX, embedded}}  - \varepsilon_n^{\text{COHSEX, gas}},
\end{align}
with $\varepsilon_n$ the n-th  energy level. The absolute energy levels for the embedded system can be obtained by adding these polarization energies to the gas phase $GW$ energy levels. In that respect, such an approach is essentially perturbative. This simple and efficient scheme will be labeled  ${\Delta}$COHSEX below.

As seen in Eqs.~\eqref{SEX} and \eqref{COH}, the COHSEX approximation to the $GW$ self-energy only requires the low-frequency limit of the screened-Coulomb potential. As such, it offers an obvious route for treating the environment in the adiabatic limit. Further, both the embedded and embedding subsystems are treated on the same footing. It remains however that the static COHSEX approximation is known to dramatically overestimate gaps, resulting from a sizeable overestimation/underestimation of the ionization potential (negative of the HOMO energy) and electronic affinity (negative of the LUMO energy). Taken as an energy difference between the gas and embedded environments, it was found that reasonable polarization energies could be obtained, stemming presumably from a cancellation of errors between the gas and embedded COHSEX calculations.  While computationally very efficient, we will show nevertheless that this approach can yield sizeable errors as compared to a fully dynamical treatment of both the embedded and embedding subsystems, presumably due to an incomplete cancellation of errors, together with the error inherent to the adiabatic approximation for the environment.

\subsubsection{ The  $\mathrm{QM}_{GW}/\mathrm{QM}_{\mathrm{COHSEX}}$   approach  }
\label{sec:gwatcohsex}

To avoid relying on the cancellation of errors at the static COHSEX level, we now merge the fully dynamical $GW$ formalism  with a polarizable environment described  in its  low-frequency susceptibility limit $\chi^{(2)}(\omega \rightarrow 0)$ .  The restriction of the correlation  self-energy to the central subsystem (1) can be obtained from equation~\eqref{eqn:sigma} by replacing $W$ by ($W_{11}-v_{11}$).  Using equations \eqref{eqn:vreac} and \eqref{eqn:tildexi} one obtains: 
\begin{align}
    \Sigma_{11}^{C}&({\bf r},{\bf r}'; E)  =   \frac{i}{2\pi}\int \dd{\omega} \mathrm{e}^{i \eta \omega} G({\bf r},{\bf r}'; E+\omega) \nonumber \\
    & \times \Big[ v_{\text{reac}}({\bf r},{\bf r}'; \omega) + [ W_{11}-\tilde{v}_{11}]({\bf r},{\bf r}'; \omega) \Big],
     \label{eqn:polarization}
\end{align}
with $G$ the subsystem (1) Green's function and
$$
[W_{11}-\tilde{v}_{11}](\omega) =  \tilde{v}_{11}(\omega) \tilde{\chi}^{(1)}(\omega) \tilde{v}_{11}(\omega),
$$
removing the space positions. The integral of the  ($W_{11}-\tilde{v}_{11}$) contribution does not present any difficulties when   $\chi^{(2)}(\omega)$ and consequently $\tilde{v}_{11}(\omega)$ are taken to their low frequency limit.  Indeed,  $\tilde{\chi}^{(1)}(\omega)$ presents a proper pole structure inherited from that of ${\chi}^{(1)}(\omega)$ through equation  \eqref{eqn:tildexi}. In the limit of a static reaction field, the poles of $\tilde{\chi}^{(1)}(\omega)$ are the ones of the gas phase ${\chi}^{(1)}(\omega)$ shifted in energy. This integral can thus be performed numerically very much as for a gas phase calculation (see Technical details in subsection \ref{Tech_Det}).

The integral involving  $v_{\text{reac}}({\bf r},{\bf r}'; \omega)$ taken in its static limit is reminiscent of the static COHSEX approximation and care must be taken to properly include the analog of the COH term.  Following the treatment exposed in Appendix \ref{part:COHSEX_pole_model},  we model the environment susceptibility as  a simple pole susceptibility $\chi^{(2)}_{\lambda}(\omega)$ with pole energy $\lambda$ (see Eq.~\eqref{eqn:ximodel}). The pole energy is taken to infinity \textit{after} performing the integration.\cite{note0} Overall, the $G v_{\text{reac}}$ contribution to the integral \eqref{eqn:polarization}, in the proper adiabatic limit,  can be reformulated as the sum of two terms
\begin{subequations}
\begin{align}
     \Sigma_{\text{vreac}}^{\text{SEX}}({\bf r},{\bf r}')    &=  - \sum_{i \in (1)}^{\mathrm{occp}} \phi_i({\bf r})\phi^*_i({\bf r}') v_{\text{reac}}({\bf r},{\bf r}'; \omega = 0)   \\   
    \Sigma_{\text{vreac}}^{\text{COH}}({\bf r},{\bf r}')    &= \frac{1}{2} \sum_{n \in (1)} \phi_n({\bf r})\phi^*_n({\bf r}') v_{\text{reac}}({\bf r},{\bf r}'; \omega=0),
\end{align}
\label{eqn:cohsexvreac}\end{subequations}
that provide the $P_n^{\,\text{COH}}$ and $P_n^{\,\text{SEX}}$ direct contributions of $v_{\textrm{reac}}$ to the polarization energies:
\begin{equation}
    P_n^{\,\text{COH\,/\,SEX}} = \langle \, \phi_n \, | \Sigma_{\text{vreac}}^{\text{COH\,/\,SEX}} | \, \phi_n \,  \rangle.
\end{equation}

As explained above, switching straightforwardly $\omega$ to zero for the reaction field in Eq.~\eqref{eqn:polarization} would lead to neglect $P^{\,\text{COH}}$. 
Anticipating on the upcoming results, we analyze in Table~\ref{tab:table1} the two contributions from Eqs.~\eqref{eqn:cohsexvreac}, together with the total polarization energy resulting from Eq.~\eqref{eqn:polarization}, in the case of a fullerene at the surface of a 302 fullerene hemisphere. The shifts in energy are presented for the HOMO, LUMO and gap of the fullerene, from the gas phase to the center of the C$_{60}$ hemisphere. We observe that while the reaction field $P^{\,\text{SEX}}$ contribution accounts for most of the polarization for the HOMO-LUMO gap, this is not the case for individual energy levels. As a matter of fact, the $P^{\,\text{SEX}}$ contribution to unoccupied states is small. Clearly, for the absolute position of individual levels, $P^{\,\text{COH}}$ cannot be neglected. A similar analysis can be found in Ref.~\citenum{Neaton2006} in the case of a molecule facing a graphene substrate. 

We note further that the $P^{\,\text{COH}}+P^{\,\text{SEX}}=P^{\,\text{COH}+\text{SEX}}$ direct contributions from the reaction field differ from the total polarization energies. The difference, reported in the last column, represents the impact of the renormalization of the integrand along the imaginary axis, namely the integral of $G [\tilde{v}{\tilde \chi}^{(1)} \tilde{v} - {v}\chi^{(1)}{v} ]$. This points to the fact that a perturbative approach where the gas phase $GW$ energy levels are only corrected by the $P^{\,\text{COH}}$ and $P^{\,\text{SEX}}$ reaction field direct contributions may lead to an error as large as about 40$\%$ considering e.g. the case of the HOMO-LUMO gap. 

 \begin{table}
\caption{\label{tab:table1}  Decomposition of the $\mathrm{QM}_{GW}/\mathrm{QM}_{\mathrm{COHSEX}}$ polarization energy (eV) for one fullerene molecule at the surface of a polarizing environment made of an hemisphere of 302 fullerenes. The $P_n^{\,\text{COH}}$ and $P_n^{\,\text{SEX}}$ terms correspond to the two contributions defined in Eqs.~\eqref{eqn:cohsexvreac}.  }
\begin{ruledtabular}
\begin{tabular}{cccccc}
            & $P_n^{\,\text{COH}}$ & $P_n^{\,\text{SEX}}$ & $P_n^{\,\text{COH+SEX}}$ &  $P_n$ & $P_n - P_n^{\,\text{COH+SEX}}$  \\
\hline
   HOMO     &  -0.98   &     1.63   &    0.65    &0.46 &  -0.19  \\ 
     LUMO     &  -1.00   &      0.26    &    -0.74    & -0.54 & 0.20 \\
   Gap       &   -0.02  &    -1.37    &    -1.39    &-1.00 &  0.39  \\

\end{tabular}
\end{ruledtabular}
\end{table}

\subsubsection{ Fully dynamically polarizable environment   }

The advantage of working with a fully \textit{ab initio} description of the environment is that the extension to dynamical screening is formally straightforward, requiring to calculate $\chi^{(2)}(\omega)$, namely the susceptibility of the environment needed to build the reaction field. Assuming that the environment is made of $N_{f}$ fragments that we label with capital indices ($\mathrm{I}=1$ to $N_{f}$), the environment susceptibility can be written:
\begin{align}
    [\chi^{(2)}]^{-1}(\omega) = \sum_{\mathrm{I}}^{N_{f}} [\chi^{(\mathrm{I})}_g]^{-1}(\omega) - \sum_{\mathrm{I} \ne \mathrm{J}}^{N_{f}} V_{\mathrm{I}\mathrm{J}}  \label{eqn:dysonfrag}
\end{align}
where $\chi^{(\mathrm{I})}_g(\omega)$ is the gas phase \textit{interacting} susceptibility of fragment (I) and where the $V_{\mathrm{I}\ne \mathrm{J}}$ are the bare Coulomb interactions coupling different fragments. Since each fragment is described by its gas phase interacting susceptibility, the Coulomb interactions inside a given fragment are already accounted for. In the $\Delta$COHSEX and $\mathrm{QM}_{GW}/\mathrm{QM}_{\mathrm{COHSEX}}$ simplified approaches, such an equation needs only to be inverted in the static $\omega=0$ limit. In the reference fully dynamical approach, the inversion is required for each real/imaginary frequency needed in our contour deformation scheme (see Technical details \ref{Tech_Det} below).  

Each isolated fragment susceptibility is constructed following Eqs.~\eqref{eqn:dysonchi} and~\eqref{eqn:xi0}. The cost of building all $\chi^{(\mathrm{I})}_g(\omega)$ grows linearly with the number $N_f$ of fragments. As a matter of fact, in the case of a C$_{60}$ crystal bulk or surface, where all fullerenes are related by translations/rotations, all $\chi^{(\mathrm{I})}_g(\omega)$ blocks can be obtained at no cost from a single fullerene gas phase susceptibility. In the limit of a very large number of fragments, the remaining cost lies essentially in inverting the Dyson equation~\eqref{eqn:dysonfrag}. We now discuss how to dramatically reduce such a cost.

\subsection{Effective polarization basis set}

In the present Coulomb fitting resolution-of-the-identity (RI-V) formulation, \cite{Ren2012,Duchemin2017b} the susceptibility for each fragment  is expressed in the auxiliary basis  set $\lbrace P \rbrace$ designed to expend the charge density and its variations:
\begin{align}
{\chi}^{(\mathrm{I})}_g(\bf{r},\bf{r'};\omega) & \;\;\overset{RI}{\simeq} & \!\!\!\!\!\sum_{P,Q} X_g^{(\mathrm{I})}(P,Q\,;\omega) \;  P({\bf{r} }) \, Q({\bf{r'}}). \label{eqn:xiref} 
\end{align} 
With the def2-TZVP-RIFIT auxiliary basis set used  in the present study, this represents 5700 basis functions per fullerene, limiting the environment to a few shells of neighbors to keep reasonable the memory and computing time associated with equation~\eqref{eqn:dysonfrag}.  

As shown in a recent publication, \cite{Amblard2023} the size of the $\chi^{(\mathrm{I})}_g(\omega)$ can be dramatically compressed when expressed in an optimal polarization basis $\lbrace \gamma \rbrace$, 
\begin{align}
{\chi}^{(\mathrm{I})}_g(\bf{r},\bf{r'};\omega) & \!\!\!\!\!\;\;\; \overset{MODEL}{\simeq} & \!\!\!\!\! \sum_{\gamma,\gamma'} \widetilde{X}_g^{(\mathrm{I})}(\gamma,\gamma'\,;\omega) \; \gamma({\bf{r} }) \, \gamma'({\bf{r'}}) \label{eqn:ximod_def}  
\end{align}
designed to preserve the resulting central fragment polarization energies within less than a meV. The effective $\lbrace \gamma \rbrace$ polarization vectors are expressed as a linear combination of the auxiliary $\lbrace P \rbrace$ basis functions. Such an optimal representation can be formulated as a variational problem, minimizing the difference between the reaction fields generated by the full and model ${\chi}^{(\mathrm{I})}_g(\omega)$ fragment susceptibilities. The components of the   $\lbrace \gamma \rbrace$     vectors and matrix coefficients $\widetilde{X}_g^{(\mathrm{I})}(\gamma,\gamma'\,;\omega)$ are the minimization parameters. Constrained minimization can be further used to preserve exactly the dipolar, quadrupolar, etc., fragment polarizabilities.    In practice, effective $\lbrace \gamma \rbrace$ polarization bases containing as little as $\sim$60 polarization vectors per C$_{60}$ can be obtained. Such a scheme allows storing and inverting with very little cost the Dyson equation~\eqref{eqn:dysonfrag} even in the limit of thousands of fullerenes in the environment. 

The construction of the effective $\lbrace \gamma \rbrace$ polarization vectors was explored and validated \cite{Amblard2023} within the framework of  the $\Delta$COHSEX scheme where only the static ${\chi}^{(\mathrm{I})}_g(\omega=0)$ are needed. We show in the Appendix \ref{part:polarbasis} that these static polarization vectors can be used to describe with preserved accuracy the dynamical 
${\chi}^{(\mathrm{I})}_g(\omega)$ operators, refitting only the $\widetilde{X}_g^{(\mathrm{I})}(\gamma,\gamma'\,;\omega)$ matrix elements for each $\omega$ (see Eq.~\eqref{eqn:ximod_def}).

We conclude this subsection by emphasizing that the central fragment and its first shell of neighbors susceptibilities are described within the full $\lbrace P \rbrace$ auxiliary basis set. Fitting by optimal polarization basis sets is thus used only for second-nearest-neighbor fragments and beyond.  

\subsection{ Technical details } \label{Tech_Det}

Our calculations are performed with the {\sc{beDeft}} (beyond-DFT) package \cite{Duchemin2020,Duchemin2021} implementing the $GW$ and Bethe-Salpeter equation (BSE) formalisms using Gaussian basis sets and Coulomb-fitting (RI-V) resolution-of-the-identity. \cite{Vahtras1993,Ren2012,Duchemin2017b}  The self-energy is calculated adopting a contour-deformation scheme for the correlation part of the self-energy : 
\begin{align*}
\Sigma^{GW}_C({\bf r},{\bf r}' ; E) &= \frac{-1}{2 \pi} \int_{-\infty}^{+\infty} \dd\omega \;   \, G({\bf r},{\bf r}' ; E+ i\omega)\, W_{\text{scr}}({\bf r},{\bf r}' ;  i\omega) \\  
-& \sum_i \phi_i({\bf r}) \phi_i^{*}({\bf r}') W_{\text{scr}}({\bf r},{\bf r}';    \varepsilon_i -E ) \theta( \varepsilon_i - E ) \\  
+& \sum_a \phi_a({\bf r}) \phi_a^{*}({\bf r}') W_{\text{scr}}({\bf r},{\bf r}'; E - \varepsilon_a) \theta( E-\varepsilon_a ) 
\end{align*}
where $(i,a)$ index occupied/unoccupied levels and  $W_{\text{scr}} = (W-v)$ the so-called screening potential. The energy integration is thus performed along the imaginary-frequency axis, completed by residues involving the value of the screened Coulomb potential along the real-axis (second and third lines).  The imaginary frequency-axis integration is performed with a 12-point quadrature optimized grid that was shown to yield quasiparticle energies at the meV-accuracy level. \cite{Duchemin2020}  Further, the screened Coulomb potential along the real-axis is obtained by analytic continuation. As shown in Ref.~\citenum{Duchemin2020}, the combination of the contour-deformation approach with the analytic continuation of the screened Coulomb potential is a very robust  scheme as compared to the direct analytic continuation of the much more structured $GW$ self-energy. Our $GW$ calculations  are performed at the non-self-consistent  $G_0W_0$ level starting with the PBE0 functional \cite{perdew-bcp-1996,adamo-jcp-1999} to generate input Kohn-Sham eigenstates with the \textsc{Orca} package. \cite{Neese2022} We adopt the def2-TZVP basis set \cite{Weigend2005} together with the corresponding  def2-TZVP-RIFIT auxiliary basis set. \cite{Weigend1998}  Since the C$_{60}$ molecule does not present any ground-state dipole, quadrupole, etc., we do not attempt to include environmental electrostatic effects in the ground-state, focusing on the dynamics of screening at the $GW$ level. 

In the case of the water-inside-nanotube system, containing 323 atoms, the tube relaxation has been performed with the \textsc{Siesta} package \cite{Soler2002} at the double-zeta plus polarization level within the local density approximation (LDA). \cite{Vosko1980} Keeping the nanotube positions frozen, the water molecule relaxation has been achieved using the  van der Waals density functional (vdW-DF) of   Dion and coworkers. \cite{Dion2004,Soler2009} Concerning our many-body calculations, since we target the response of a metallic tube, we favored input Kohn-Sham eigenstates generated at the def2-TZVP LDA level with the \textsc{Orca} package.
For sake of efficiency, the nanotube susceptibility has been calculated within our recently implemented cubic-scaling real-space imaginary-time approach \cite{Duchemin2019,Duchemin2021} inducing errors at the meV level for quasiparticle energies as compared to a def2-TZVP/def2-TZVP-RIFIT calculation within our standard Coulomb-fitting resolution-of-identity implementation. The fully dynamical $GW$ calculation for the water-plus-nanotube system, involving the calculation of the nanotube $\chi^{(2)}(i\omega)$  for 12 imaginary frequencies, required no more than a thousand CPU hours in total at the def2-TZVP level. \cite{cpunote}

\section{ Results and discussions }

\begin{figure}
  \includegraphics[width=14cm]{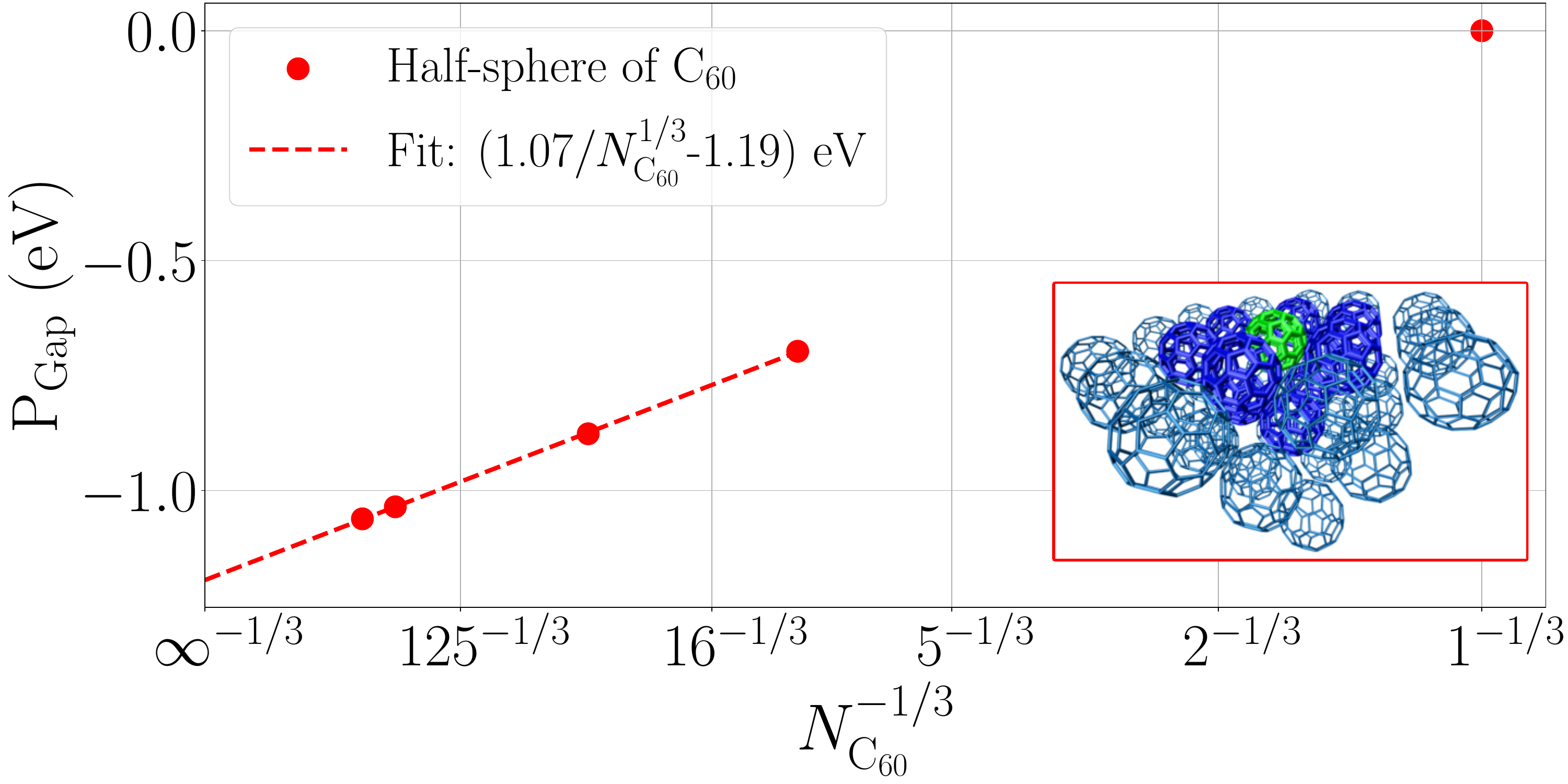} 
  \caption{ Evolution of the gap polarization energy for a  C$_{60}$ located at  the surface of a FCC crystal hemisphere containing $N_{\mathrm{C}{60}}$ fullerenes (see Inset with in green the ``central'' surface fullerene). The dashed line is the linear fit as a function of $N_{\mathrm{C}{60}}^{-1/3}$ which is proportional to $1/R$, with $R$ the hemisphere radius. The red dots points to systems with $N_{\mathrm{C}{60}}$=1, 10, 37, 302 and 534. }
  \label{fig:fulldyn}
\end{figure}

We now study the case of a fullerene at the (111) surface of a face-centered cubic (FCC) C$_{60}$ crystal. \rep{\strikeout{In the case of organic systems}}  Photoemission spectroscopy is very much surface sensitive due to the large \rep{\strikeout{adsorption}} \rep{absorption} of incoming photons/electrons by the first layer.  Fullerene surfaces have been extensively studied experimentally, \cite{Reihl1994,Weaver1992,Benning1992,Lof1992,Takahashi1992}  providing valuable reference data.
We plot in Fig.~\ref{fig:fulldyn}  the polarization energy associated with the gap of a surface C$_{60}$ as a function of the number $N_{\mathrm{C}{60}}$ of fullerenes in a surrounding hemisphere (see Inset).  In the fragment approximation, without the formation of bands originating from wavefunction overlap, it is the so-called peak-to-peak gap that is studied, that is the energy difference between the centers of the HOMO and LUMO bands. 
Calculations are performed at the fully dynamical level without any adiabatic (instantaneous response) approximation for the environment. This will serve as a reference for calculations performed in the static susceptibility limit for the surrounding fullerenes.

Due to  screening, the surface fullerene gap is closing with a $(1/R)$ scaling law,  
 with $R$ the radius of the environment hemisphere. Equivalently, the gap closes linearly with respect to $N_{\mathrm{C}{60}}^{-1/3}$, where $N_{\mathrm{C}{60}}$ is the total number of fullerenes.  The quality of the linear fit indicates that  the considered $N_{\mathrm{C}{60}}$ are large enough to enter the asymptotic $1/R$ scaling law regime. In the infinite limit, our dynamical calculations provide a polarization energy of -1.2 eV for the surface fullerene gap. This can be compared with experimental values of -1.1 eV,\cite{Reihl1994} -1.2 eV \cite{Weaver1992,Benning1992} or -1.4 eV  \cite{Lof1992,Takahashi1992}  obtained by subtracting the experimental 4.9 eV gas phase fullerene HOMO-LUMO gap \cite{NIST} to the experimental surface peak-to-peak  gap.    
 We show in the Supplementary Material  (SM) how to recover the same surface limit by growing the crystal layer-by-layer, as in a slab calculation, with a   $(1/n)$   convergence of the polarization energy, where  $(n)$  is the number of layers (Fig.~S1).

\begin{figure}
  \includegraphics[width=14cm]{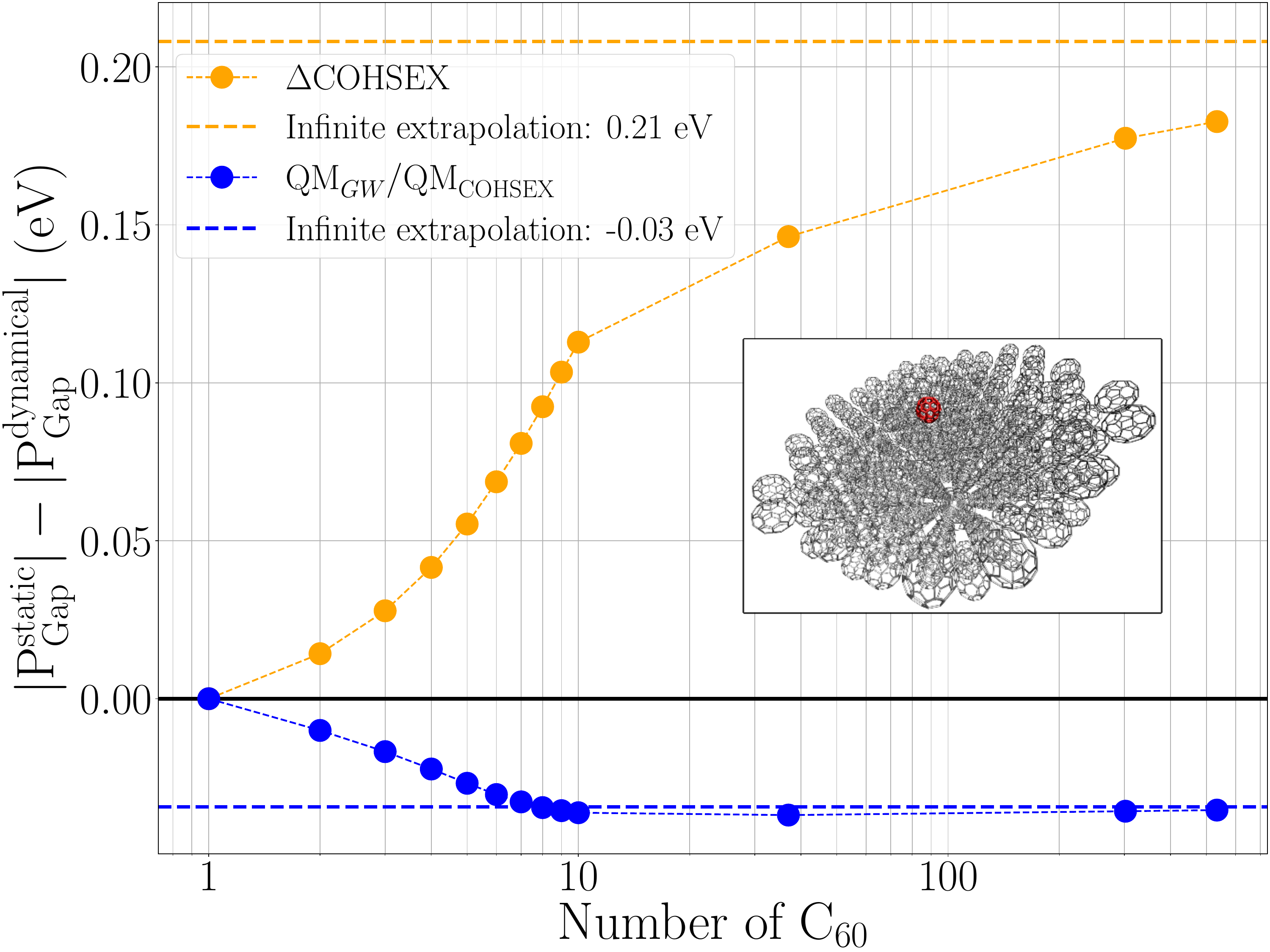} 
  \caption{ Errors, in eV, on the gap polarization energy when using the reaction field in its static limit within the $\Delta$COHSEX (orange dots) and $\mathrm{QM}_{GW}/\mathrm{QM}_{\mathrm{COHSEX}}$ (blue dots) schemes, as compared to a fully dynamical calculation. 
  Error calculated as a function of the number of fullerenes. 
  The first and second shells of neighbors correspond to $N_{\mathrm{C}{60}}$= 10 and 37.
  Abscissa axis in log-scale. Dashed lines represent the errors in the extrapolated infinite size limit. Inset: System with 302 C$_{60}$. The ``central'' surface fullerene is in red. }
  \label{fig:errors}
\end{figure}

We now address the central issue of the present study, namely the impact of assuming that the environment responds instantaneously to an electronic excitation in a surface fullerene. In the case of a fullerene inside a fullerene crystal, the decoupling of excitation energies between the central subsystem and its environment is clearly not satisfied. The resulting errors on the polarization energy, as compared to a fully dynamical calculations,  are provided in Fig.~\ref{fig:errors}. 

Considering first  the case of static $\Delta$COHSEX calculations (orange dots), we observe that the error on the polarization energy grows with the number of surrounding fullerenes. As a matter of fact, the error grows rapidly when completing the first shell of neighbors (up to $N_{\mathrm{C}{60}}=10$), more slowly when completing the second shell of neighbors (up to $N_{\mathrm{C}{60}}=37$), etc. Asymptotically, the total error on the polarization energy amounts to 0.208 eV, namely a very sizeable 17.5$\%$ of error as compared to the fully dynamical calculation. 
A positive error means that the $\Delta$COHSEX scheme overestimates the closing of the gap by screening. 

We now address the case of the $\mathrm{QM}_{GW}/\mathrm{QM}_{\mathrm{COHSEX}}$ scheme (blue dots), merging a fully dynamical $GW$ calculation for the central subsystem with the proper static screening limit for the environment. The error is found to be significantly reduced as compared to the $\Delta$COHSEX scheme, with an asymptotic error of -0.034 eV, amounting to a much reduced 2.9$\%$ error
with respect to the fully dynamical asymptotic $P_{\mathrm{Gap}}$ = -1.19 eV. As another important difference with the $\Delta$COHSEX scheme, the error essentially builds within the first-shell of neighbors. This really means that in the $\mathrm{QM}_{GW}/\mathrm{QM}_{\mathrm{COHSEX}}$ scheme, only the closely lying molecular fragments really need to be treated at the full dynamical level to fully reproduce the effect of the environment dynamical response. \textcolor{black}{Similar results can be obtained for the polarization energy associated with the individual HOMO and LUMO levels (SM, Fig.~S2). Analogous plots, but with an error given  in percentage, can also be found in the SM  (Fig.~S3). }

The reason why the static approximation induces larger errors at short-range, namely for polarizable fragments located close to the central subsystem of interest, was  hinted in Ref.~\citenum{Neaton2006}. Focusing e.g. on the $\phi_H$ HOMO eigenstate with energy $E_H$, the fully-dynamical SEX-like contribution  to the polarization energy  reads: 
\begin{align*}
   \langle \phi_H | \Sigma_{\text{vreac}}^{\text{SEX}}(E_H) | \phi_H \rangle &= -\sum_{i \in (1)}^{\text{occp}} \langle \phi_H \phi_i | v_{\text{reac}}(E_H-E_i) | \phi_i \phi_H \rangle  .
\end{align*}
 In the limit of a smoothly varying reaction field over the extent of the central subsystem, orthogonalization of the molecular orbitals reduces the sum to the ($i$=HOMO) terms, and only the static $v_{\text{reac}}(E_H-E_H=0)$ contribution is required. The same reasoning holds for the $P^{\,\text{COH}}$ term.  Adding far standing fullerenes amounts to adding components of the reaction field that are more and more smoothly varying on the central subsystem, with decreasing contribution to the total error associated with the static limit.  In the same smoothly varying limit, one observes that the $P^{\,\text{SEX}}$ contribution vanishes for the polarization energy associated with unoccupied states, while for occupied states the $P^{\,\text{COH}}$ term amounts to (-1/2) of the $P^{\,\text{SEX}}$ contribution. \cite{Neaton2006} 

To further explore the impact of the adiabatic limit for the polarizable medium, we  consider an environment of fictitious fullerenes where the HOMO-LUMO gap has been artificially changed in a scissor fashion, namely moving rigidly in energy the occupied manifold with respect to the unoccupied one. Such modified Kohn-Sham energy levels are used to construct the $\chi^{(2)}(\omega)$ susceptibility of the environment in the fully dynamical, $\Delta$COHSEX and $\mathrm{QM}_{GW}/\mathrm{QM}_{\mathrm{COHSEX}}$ limits. The PBE0 energy levels for the central (surface) C$_{60}$ are not modified (PBE0 HOMO-LUMO gap of 2.99 eV). 

\begin{figure}
  \includegraphics[width=14cm]{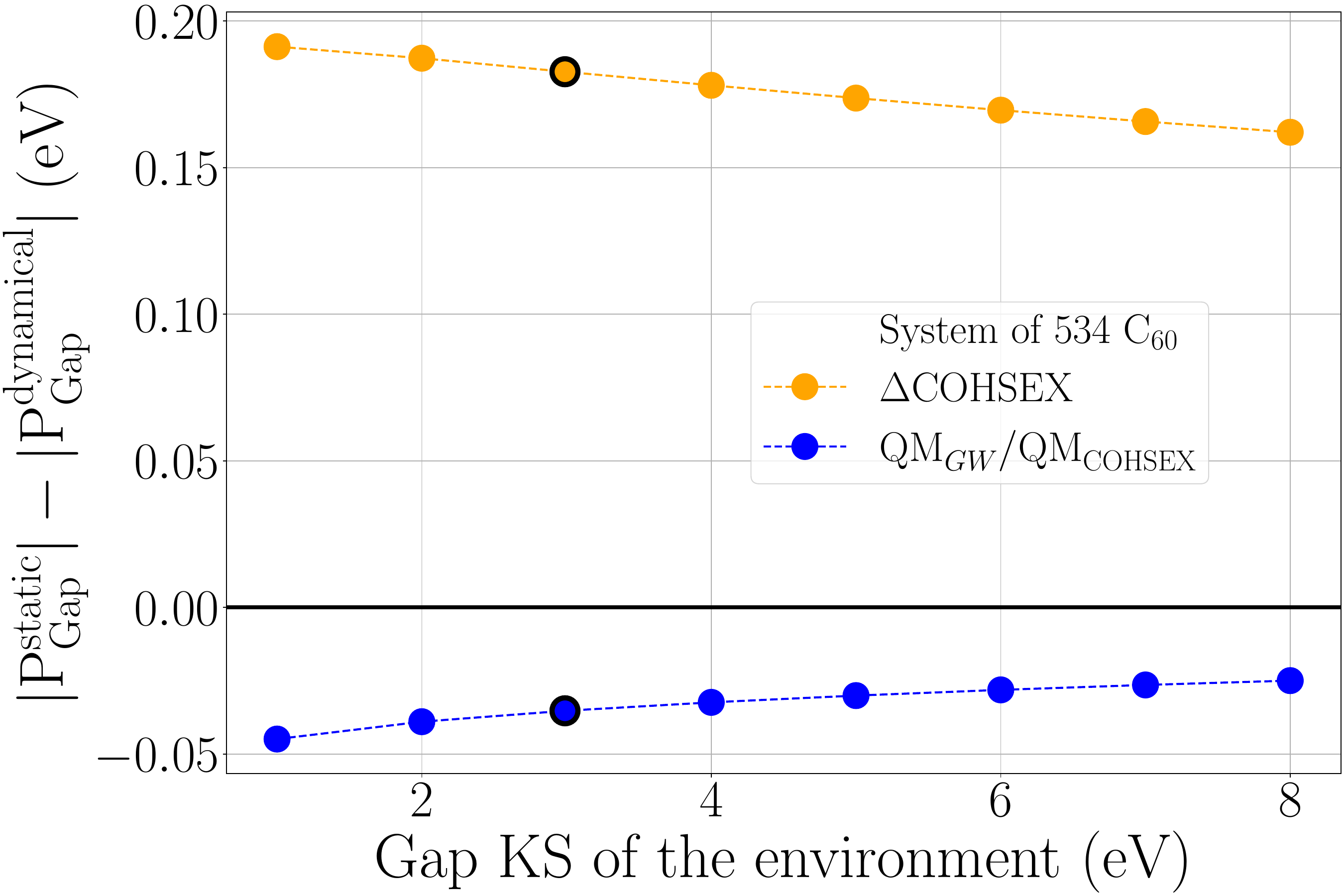} 
  \caption{ Errors, in eV, on the gap polarization energy when using the reaction field in its static limit as compared to a fully dynamical calculation. The error is calculated as a function of the fictitious gap of the surrounding fullerenes (see text). The color code is the same as in Fig.~\ref{fig:errors}. Data with black bold edges correspond to the true Kohn-Sham (PBE0) fullerene gap ($2.99$ eV).   }
  \label{fig:gapsfictifs}
\end{figure}

We plot in Fig.~\ref{fig:gapsfictifs} the error as a function of the modified gap, keeping constant here the number of surrounding fullerenes ($N_{\mathrm{C}{60}}=534$).  As expected, the error increases/decreases when the input HOMO-LUMO gap of the surrounding fullerenes is decreased/increased with respect to the unmodified central surface fullerene. When the gap of the surrounding fullerenes becomes larger, the adiabatic approximation for the environment becomes formally better validated and the  error induced by the static approximation is reduced. Again, both in the small and large environment gap limits, the $\mathrm{QM}_{GW}/\mathrm{QM}_{\mathrm{COHSEX}}$ scheme (blue dots) provides the smallest error, with a  3.9$\%$  maximum error in the \rep{\strikeout{antiadiabatic limit for the environment}} limit of a small environmental gap.   \rep{In such a limit, the adiabatic approximation for the surrounding medium is expected to fail. We provide in the SM (Fig.~S4) the same plot but for the individual HOMO and LUMO levels, indicating that there is no compensation between the errors affecting the HOMO and LUMO polarization energies, respectively.  }

We conclude this exploration by considering the case of a water molecule inserted inside a long section of a metallic (10,10) carbon nanotube. The behavior of water inside carbon nanotubes has been the subject of many studies, targeting a better understanding of the structure of confined water and the nature of   friction at the water/tube interface \cite{Falk2010} for applications to energy generation through inverse osmosis and water desalinization.   Such studies stand much ahead of the present exploration aiming at better understanding the impact of dynamical versus static screening. The case of a metallic  surrounding medium  offers a stringent test for the assumption that the environment responds instantaneously to an electronic excitation in the central subsystem. 

\begin{figure}
   \includegraphics[width=4.6cm]{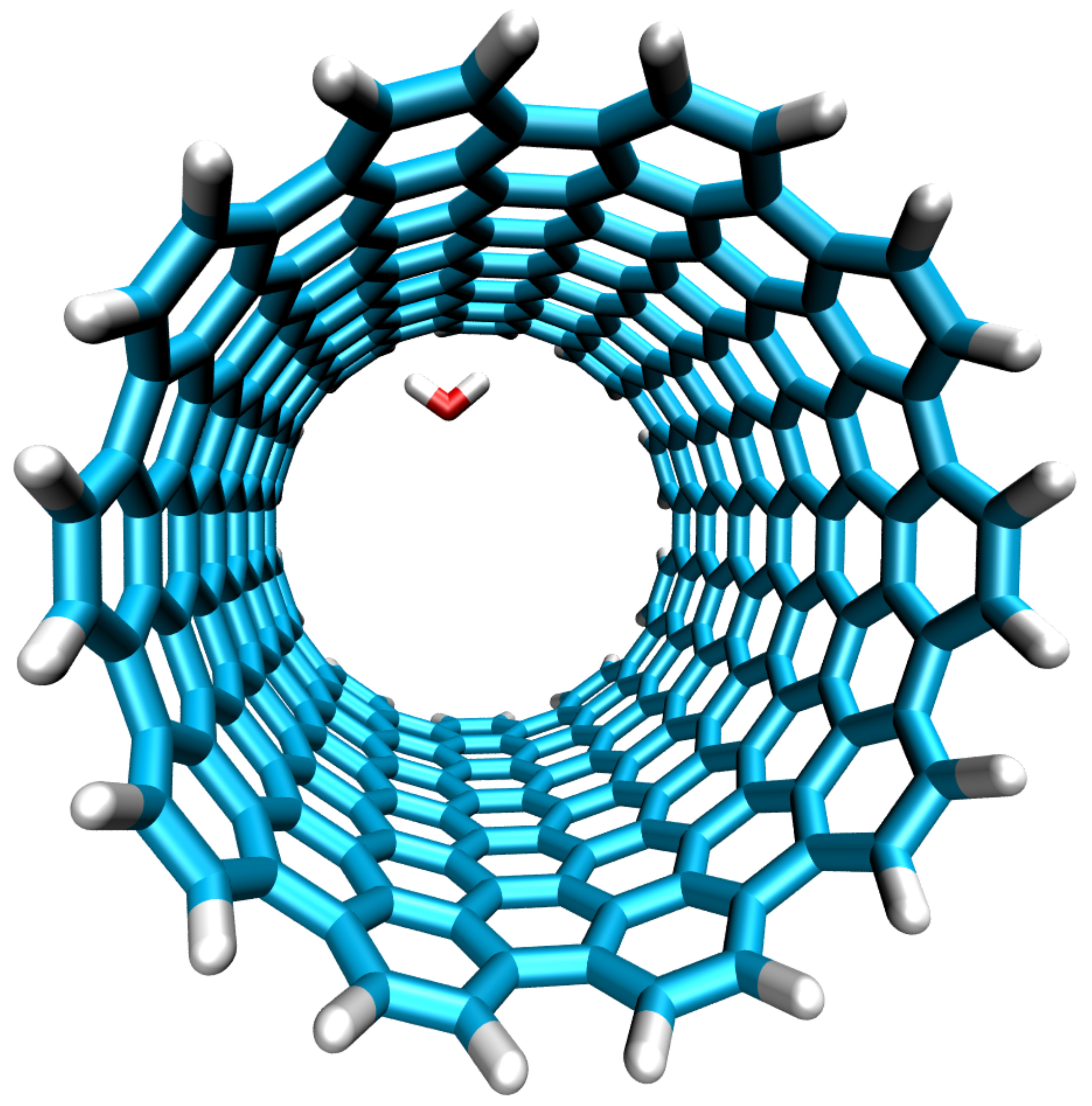} 
  \caption{ Geometry of the H$_2$O@(10,10) nanotube section model system.   The nanotube section contains 280 carbon atoms with hydrogen-passivated edges.   The water molecule adopts a close to ``{o}ne-leg'' geometry \cite{Ma2011} with an OH bond pointing towards a hexagon center.   }
  \label{fig:watertube}
\end{figure}

Our model system is represented in Fig.~\ref{fig:watertube}.  Since we are interested in the electronic response properties of a (nearly) metallic medium, we here favour the  local density approximation  (LDA)   over hybrid functionals. Due to finite size, the LDA  gap is finite, amounting to 0.3   eV. Even though not strictly metallic, the nanotube gap is significantly smaller than that of the water molecule (6.92 eV, LDA value).  The water and nanotube are treated as separate fragments, namely we do not allow hybridization between them and focus on long-range screening. The closing of the water molecule HOMO-LUMO gap from the gas phase to the nanotube-intercalated geometry amounts to -2.09 eV, -2.23 eV and -2.41 eV,  at the fully dynamical, $\mathrm{QM}_{GW}/\mathrm{QM}_\mathrm{COHSEX}$ and $\Delta$COHSEX schemes, respectively. Again the $\mathrm{QM}_{GW}/\mathrm{QM}_\mathrm{COHSEX}$ scheme yields the smallest error ($\sim$7$\%$)  as compared to the $\Delta$COHSEX approach ($\sim$15$\%$ error). \rep{ This is consistent with the data of Fig.~\ref{fig:gapsfictifs} in the limit of a small gap environment. }
\rep{ The error on the gap polarization energy benefits however here from a small compensation of errors. At the ${\Delta}$COHSEX and $\mathrm{QM}_{GW}/\mathrm{QM}_\mathrm{COHSEX}$ levels, the $P_{\text{HOMO}}$ increases by $\sim 360$ meV and $\sim 225$ meV  respectively, as compared to the fully dynamical calculation, a rather large variation. However, the $P_{\text{LUMO}}$ reduces by $\sim 40$ meV and $\sim 90$ meV, respectively, in absolute value.  }

As a conclusion, we observe that even in a situation where the  environment adiabatic treatment is expected to fail, the static reaction field approximation yields an error on the gap smaller than 10$\%$ when treated correctly. \rep{This is consistent with the findings of Ref.~\citenum{Neaton2006} where the case of a benzene molecule facing a graphene sheet was considered. We note that what should matter is the relative position of the neutral excitation spectra for the embedded and embedding subsystems. In that respect, comparing just the HOMO-LUMO gaps may not stand as an accurate criteria. }

\section{ Conclusions }

Within the framework of an embedded QM/QM' $GW$  framework, we have studied the validity of  the adiabatic approximation for the environment, namely the assumption that the electronic degrees of freedom in the environment (treated at the QM' level) respond instantaneously to an excitation in the central subsystem (treated at the QM level). In practice, this amounts to restricting the environment electronic susceptibility to its low frequency $\chi^{(2)}(\omega \rightarrow 0)$ limit, where (2) points to the environment. As a first test case, we have in particular explored a fullerene at the surface of a fullerene crystal, a paradigmatic organic crystal where there is no decoupling in the excitation energy spectrum  between the embedded and embedding subsystems. Further, we have studied the case of a water molecule inside a metallic tube section regarded as the environment. In such a situation, the adiabatic limit for the environment is formally not expected to be valid.

Reference  calculations are performed at the $\mathrm{QM}_{GW}/\mathrm{QM}_{GW}$  level, where both embedded and embedding subsystems are treated at the fully dynamical level within a fragment approximation. Our findings are that the proper adiabatic limit for the environment consists in a $\mathrm{QM}_{GW}/\mathrm{QM}_\mathrm{COHSEX}$ approach where the reaction field from the environment is treated in the same fashion as the static COHSEX (Coulomb-hole plus screened exchange) approximation to the full $GW$ self-energy. Maximum errors on the polarization energy, that is the energy shift of  the energy levels from the isolated (gas) to the embedded  geometry, are found to remain below 10$\%$, \rep{except for the case of a metallic environment (nanotube) where the shift on the individual frontier orbitals can be as large as a few tenths of an eV}. If better accuracy is desired, we found that treating the first-nearest-neighbors only at the fully dynamical level, while assuming the adiabatic limit for farther lying fragments, allows to reduce the error to negligible values. Such a  scheme may be regarded as a  $\mathrm{QM}_{GW}/\mathrm{QM}_{GW}/\mathrm{QM}_\mathrm{COHSEX}$  approach. 

In contrast, the scheme employed in previous work consisting in following the shift of the energy levels from a gas phase $\mathrm{QM}_\mathrm{COHSEX}$ to an embedded $\mathrm{QM}_\mathrm{COHSEX}/\mathrm{QM}_\mathrm{COHSEX}$ calculation, namely an approach where both the embedded and embedding subsystems are treated consistently in the static COHSEX limit, is found to induce much larger errors.  As compared to the more accurate $\mathrm{QM}_{GW}/\mathrm{QM}_\mathrm{COHSEX}$ calculation, we speculate that the well-known COHSEX inaccuracy, as compared to $GW$,  for the calculation of the energy levels of the central (embedded) subsystem, does not fully cancel out between the gas and embedded systems. 

\rep{ The fragment approximation transfers most of the computer time requirements to the cubic-scaling inversion of the Dyson equation for the screened Coulomb potential. In the fully dynamical scheme, this inversion must be performed $n{\omega}$ times, with $n{\omega}$ the number of frequencies sampling the real or imaginary axis ($n{\omega}$=12 in the present study~\cite{Duchemin2020}). In the   $\mathrm{QM}_{GW}/\mathrm{QM}_\mathrm{COHSEX}$ scheme, the construction of the static $v_{\text{reac}}(\omega=0)$ reaction field requires   inverting only once a Dyson equation over the environment degrees of freedom. In the limit of large systems, the dynamical scheme is thus formally $n{\omega}$ times more expansive than the $\mathrm{QM}_{GW}/\mathrm{QM}_\mathrm{COHSEX}$ one. In practice, due to the fixed cost of calculating the bare Coulomb potential matrix elements over the environment within the RI-V approach, an embedded $GW$ calculation over 534 fullerenes amounts, in the present stage of implementation, to 420 CPU hours and 150 CPU hours  at the $\mathrm{QM}_{GW}/\mathrm{QM}_{GW}$   and $\mathrm{QM}_{GW}/\mathrm{QM}_\mathrm{COHSEX}$ levels, respectively (def2-TZVP non-self-consistent $G_0W_0$ calculations). }

The present results, concerned with the treatment of the electronic dynamics, is expected to remain valid for approaches attempting to go beyond the fragment approximation. Strategies  based on  fragments-dimer corrections, \cite{Fujita2021}  or corrections relying on calculating the  susceptibility of fully interacting fragments but restricted to a small subset of occupied/unoccupied states, \cite{Cheng2021,Li2023} can be fully combined with the present scheme. Similarly, the same results should hold for polarizable continuum models (PCM) of environments usually treated by considering only the  optical dielectric response (square of the refractive index) in its low frequency limit, the so-called $\epsilon_{\infty}$ constant (e.g. 1.78 for water).

 \section*{ SUPPLEMENTARY MATERIAL }

{ See the Supplementary Material for (a) the convergence to the surface polarization energy growing the fullerene crystal layer-by-layer, (b) the equivalent of Fig.~\ref{fig:errors}  but for the HOMO and LUMO energy levels, (c)  in percentage  rather than in absolute values, \rep{and (d) the equivalent of Fig.~\ref{fig:gapsfictifs} but for the individual HOMO and LUMO levels.}   }

\begin{acknowledgments} 
 XB acknowledges numerous discussions with Gabriele D'Avino.
 DA is indebted to ENS Paris-Saclay for his PhD fellowship. 
 This project was provided with computer and storage ressources by GENCI@TGCC thanks to the grants A0130910016 and A0150910016 on the  Joliot-Curie supercomputer (SKL and Rome partitions).
 XB and ID acknowledge support from the French Agence Nationale de la Recherche (ANR) under contract ANR-20-CE29-0005.
\end{acknowledgments} 

 \section*{Data Availability Statement}

The data that supports the findings of this study are available within the article and its associated Supplementary Material.  


\appendix
\section{Static COHSEX from a simple pole model}\label{part:COHSEX_pole_model} 

We recover here the static COHSEX approximation within a very simple pole model for the susceptibility (see Eq.~\eqref{eqn:ximodel}) as an alternative to original approaches relying on the time-domain analysis of the response operators. \cite{Hed65,Bruneval2005,Bruneval2006} 
The static COHSEX approximation can be recovered by shifting the pole energy to infinity.  
Separating the self-energy $\Sigma=\Sigma^{\text{X}}+\Sigma^{\text{C}}$ into its bare-exchange part $\Sigma^{\text{X}}$ and its correlation part $\Sigma^{\text{C}}$, with
\begin{align}
    \Sigma^{\text{X}}({\bf r},{\bf r}')&= \frac{i}{2\pi} \int \dd\omega \; \mathrm{e}^{i \eta \omega} G({\bf r},{\bf r}'; E+\omega)  v({\bf r},{\bf r}') \\
    &=- \sum_i^{\text{occp}}  \phi_i({\bf r}) \,\phi^{*}_i({\bf r}') \, v({\bf r},{\bf r}'),
\end{align}
the correlation-only self-energy reads :
\begin{align}
     \Sigma^{\text{C}}_\lambda({\bf r},{\bf r}';E)&= \frac{i}{2\pi} \int \dd\omega \dd{\bf r}_1\dd{\bf r}_2\; \mathrm{e}^{i \eta \omega} G({\bf r},{\bf r}'; E+\omega)  \\
     &\times v({\bf r},{\bf r}_1)\chi_\lambda({\bf r}_1,{\bf r}_2; \omega) v({\bf r}_2,{\bf r}'). \nonumber
\end{align}
Using Eq.~\eqref{eqn:ximodel} and the two following integral identities
\begin{align}
    &\int \dd{\omega}\frac{1}{\omega+E-(\varepsilon_i+i\eta)} \left(\frac{1}{\omega+\lambda-i\eta}-\frac{1}{\omega-\lambda+i\eta}\right)  \\
    & = \phantom{-}\frac{2i\pi}{\lambda+E-\varepsilon_i} \nonumber \\
    &\int \dd{\omega}\frac{1}{\omega+E-(\varepsilon_a-i\eta)} \left(\frac{1}{\omega+\lambda-i\eta}-\frac{1}{\omega-\lambda+i\eta}\right)  \\
    &= -\frac{2i\pi}{\lambda+\varepsilon_a-E}, \nonumber 
\end{align}
leads to 
\begin{align}
     \Sigma^{\text{C}}_\lambda({\bf r},{\bf r}';E)&= -\int \dd{\bf r}_1 \dd{\bf r}_2
     v({\bf r},{\bf r}_1) {\chi}({\bf r}_1,{\bf r}_2; 0)v({\bf r}_2,{\bf r'})  \\
     &\times \frac{\lambda}{2}  \left[ \sum_i \frac{\phi_i({\bf r})\phi^*_i({\bf r}')}{\lambda+E-\varepsilon_i} - \sum_a \frac{\phi_a({\bf r})\phi^*_a({\bf r}')}{\lambda+\varepsilon_a-E} \right] \nonumber
\end{align}
with $(i)$ indexing   occupied states, and $(a)$ the empty ones. The limit $\lambda \to \infty$ results in the frequency-independent correlation part $\Sigma^{\text{C}}_\infty$ such that
\begin{align}
    \Sigma^{\text{C}}_\infty({\bf r},{\bf r}') &= \frac{1}{2} \bqty{W({\bf r},{\bf r}'; \omega=0) - v({\bf r},{\bf r}')} \label{eqn:SigmaC_COHSEX}\\
    &\times \left[ \sum_a  \phi_a({\bf r})\phi^*_a({\bf r}') - \sum_i  \phi_i({\bf r})\phi^*_i({\bf r}') \right],  \nonumber
\end{align}
where the sums $[\sum_a - \sum_i ]$ can be reformulated as $[\sum_n - 2\sum_i ]$, leading to the static COHSEX approximation : 
\begin{align}
    \Sigma({\bf r},{\bf r}')&=\Sigma^{\text{X}}({\bf r},{\bf r}')+ \Sigma^{\text{C}}_\infty({\bf r},{\bf r}') \\
    &=\Sigma^{\text{SEX}}({\bf r},{\bf r}') + \Sigma^{\text{COH}}({\bf r},{\bf r}'),
\end{align}
as defined in Eqs.~\eqref{SEX} and ~\eqref{COH}. We have used the same approach in Sec.~\ref{sec:gwatcohsex} to obtain the proper expression for the contribution of the reaction field in the static limit for the environment $\chi^{(2)}$ susceptibility. 

\section{ Effective polarization basis for dynamical susceptibilities }\label{part:polarbasis} 

In a previous study, \cite{Amblard2023} we have shown that the static interacting susceptibility $\chi^{(\mathrm{I})}({\bf r},{\bf r}'; \omega=0)$ associated with a given fragment (I) in the environment could be expressed in a minimal ``polarization basis'' $\lbrace \gamma \rbrace$ while preserving the reaction field on the central subsystem with an error at the meV level on the polarization energies. Following the definition of the susceptibility expressed in the full auxiliary basis $\lbrace P \rbrace$ (Eq.~\eqref{eqn:xiref}), and the model susceptibility  expressed in the  optimal polarization basis (Eq.~\eqref{eqn:ximod_def}), the optimal vectors $\lbrace \gamma \rbrace$ and corresponding matrix elements $\widetilde{X}_g^{(\mathrm{I})}(\gamma,\gamma'\,;\omega)$ can be obtained through a minimization problem: 
\begin{equation}
\begin{split}
\argmin_{\{\gamma\} } \left(\;\min_{\big\{\widetilde{X}_g^{(\mathrm{I})}(\gamma,\gamma'\,;\omega)\big\}} 
\sum_{t,t'} \left| \langle t\, | \Delta v_{\text{screen}}^{(\mathrm{I})}(\omega) |\, t' \rangle \right|^2  \;\right),
 \label{eqn:ximod_dirs}
\end{split}
\end{equation}
where $\lbrace t \rbrace$ are test functions including the auxiliary basis $\lbrace P \rbrace$ augmented with very diffuse orbitals allowing to test the quality of the model reaction field in the vicinity of the fragment. In this minimization process, the difference $\Delta v_{\text{screen}}^{(\mathrm{I})}$ reads  
\begin{equation}
\begin{split}
& \Delta v_{\text{screen}}^{(\mathrm{I})}({\bf r},{\bf r'};\omega) \phantom{\Bigg)}\\
& =  \iint \dd{\bf r}_1 \dd{\bf r}_2 \; v({\bf r},{\bf r}_1)\  
\Delta \chi_{g}^{(\mathrm{I})}({\bf r}_1,{\bf r}_2;\omega)\  v({\bf r}_2,{\bf r}').
 \label{eqn:dvreac}
\end{split}
\end{equation}
with
\begin{equation}
\begin{split}
\Delta \chi_{g}^{(\mathrm{I})}({\bf r},{ \bf r'};\omega) & =
\sum_{\gamma,\gamma'} \widetilde{X}_g^{(\mathrm{I})}(\gamma,\gamma'\,;\omega) \; \gamma({\bf{r} }) \, \gamma'({\bf{r'}}) \\
& -\sum_{P,Q} X_g^{(\mathrm{I})}(P,Q\,;\omega) \;  P({\bf{r} }) \, Q({\bf{r'}})
 \label{eqn:dximod}
\end{split}
\end{equation}
The minimization process can be performed under the constraint that the fragment polarizability tensor is preserved by the model susceptibility, insuring the proper long-range reaction field. All details can be found in Ref.~\citenum{Amblard2023} in the static $(\omega \rightarrow 0)$ limit. 

In the present study, the same scheme was adopted for the needed dynamical susceptibilities  $\widetilde{X}_g^{(\mathrm{I})}(\gamma,\gamma'\,;\omega)$. For sake of simplicity, we adopt the static optimal polarization basis vectors $\lbrace \gamma(\omega=0) \rbrace$ and only reoptimize the matrix elements $\widetilde{X}_g^{(\mathrm{I})}(\gamma,\gamma'\,;\omega) $ for each frequency. A similar choice was made in pioneering studies involving generalized plasmon-pole models \cite{Rohlfing1995} or  effective representation of dielectric matrices \cite{Wilson2008,Govoni2015,DelBen2019}  expressing the dynamical susceptibility in a basis of polarization vectors obtained as the leading eigenvectors of a symmetrized   $(\sqrt{v} \cdot \chi(\omega=0) \cdot \sqrt{v})$ static susceptibility, using symbolic notations.

\begin{figure}
   \includegraphics[width=8.6cm]{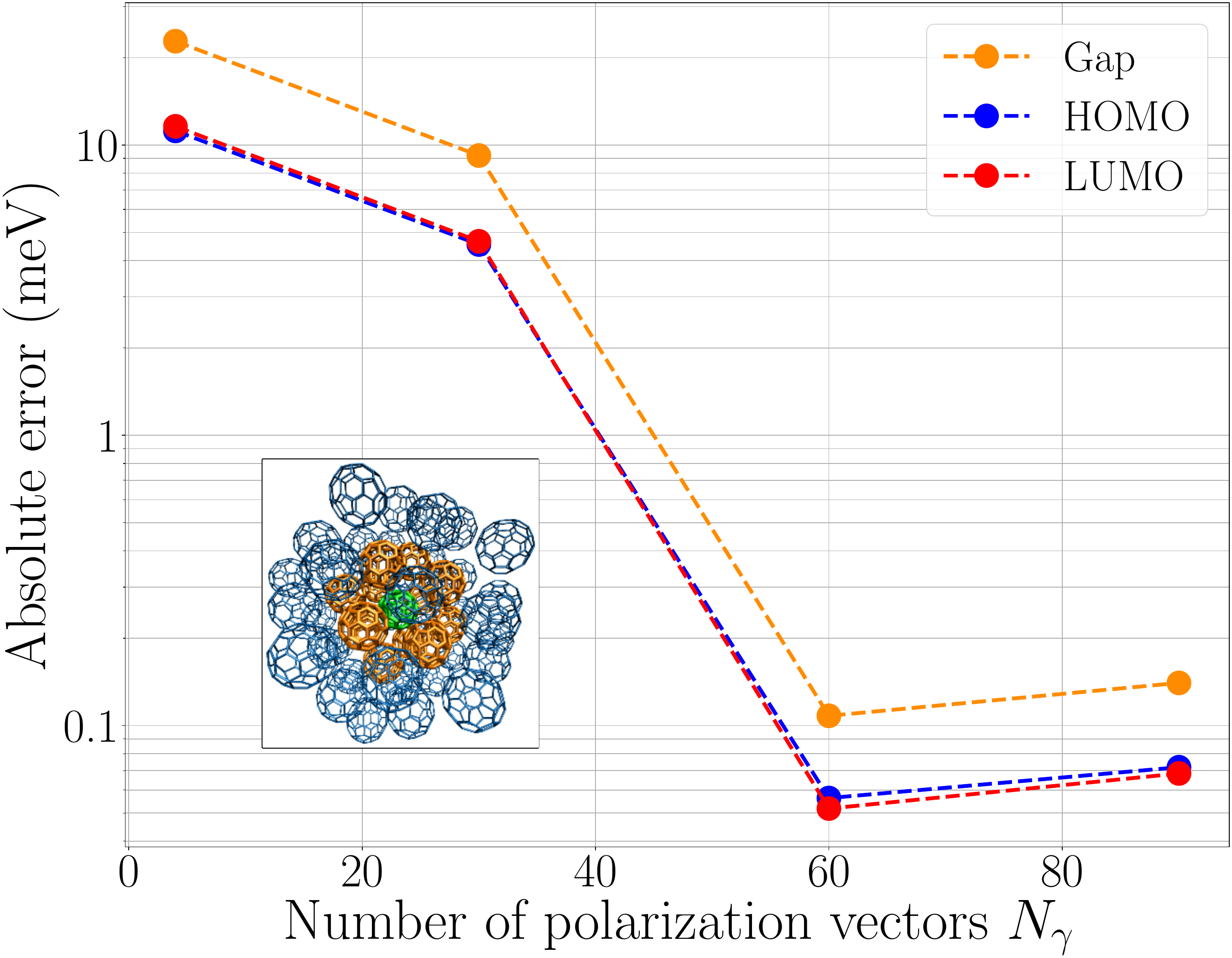} 
  \caption{  Absolute value of the errors on the gap, HOMO and LUMO energy levels, for a fullerene surrounded by its two first-shells of neighbors in a sphere (bulk) geometry. The susceptibility of the central (in green, Inset) and 12 first-nearest-neighbors (in orange, Inset) C$_{60}$ are described by the full auxiliary basis (5700 orbitals), while the susceptibility for each of the 42 C$_{60}$  in the second shell of neighbors (in blue, Inset) is described by $N_{\gamma}$ polarization vectors. Energies on the ordinates are in meV and log-scale. Errors for the HOMO are negative. Calculations performed at the full dynamical level both for the embedded and embedding subsystems.  }
  \label{fig:errorfit}
\end{figure}

To show the accuracy of this approach, we study the energy levels (HOMO, LUMO and gap) for a central fullerene surrounded by two shells of neighbors (see Inset Fig.~\ref{fig:errorfit}). We plot the error associated with replacing the full susceptibility operator by the model susceptibility beyond the first-shell of neighbors. The error is plotted as a function of the number $N_\gamma$ of chosen static $\lbrace \gamma \rbrace$ vectors per fullerene. The error  goes below the meV for a number of $\lbrace \gamma \rbrace$ functions as small as 60, namely no more than the number of atoms in a fullerene. This can be compared to 5700, the size of the full def2-TZVP-RIFIT auxiliary basis set per fullerene. Due to the strict preservation of the fullerene polarizability tensor in the constrained fitting process, the error induced using a model susceptibility with $N_{\gamma}=60$ for fullerenes located beyond the second nearest neighbors becomes negligible. This allows to invert  the Dyson equation (Eq.~\eqref{eqn:dysonfrag}) at each frequency for an environment involving hundreds of fullerenes with very limited CPU and memory requirements. 
\bibliography{xavbib.bib}

\begin{thebibliography}{138}
\bibitem [{not()}]{note0}%
  \BibitemOpen
  \href@noop {} {}\bibinfo {note} {For sake of consistency, one can verify that
  using a one-pole $\chi^{(2)}_{\lambda}(\omega)$ dynamical model to evaluate
  the integral involving $( W_{11}-\tilde{v}_{11})( \omega)$, and taking
  $\lambda$ to infinity after performing the integration, allows to recover the
  same result as taking directly $v_{\mathrm{reac}}(\omega)$ to its static
  limit in $( W_{11}-\tilde{v}_{11})( \omega)$ as discussed in the main
  text.}\BibitemShut {Stop}%
\end{thebibliography}

\end{document}